\documentclass[runningheads]{llncs}
\usepackage{graphicx}
\usepackage{cite}
\usepackage{amsmath,amssymb,amsfonts}
\usepackage{algorithmic}
\usepackage{graphicx}
\usepackage{textcomp}
\usepackage{xcolor}
\usepackage[T1]{fontenc}
\usepackage{longtable}
\usepackage{booktabs}
\usepackage{array}
\usepackage{tabularx}
\usepackage{adjustbox}
\usepackage{framed}
\usepackage[multiple]{footmisc} 
\usepackage{ragged2e}
\usepackage{xurl}
\usepackage[colorlinks=true,urlcolor=blue]{hyperref}
\usepackage{multirow}
\usepackage{hanging}
\usepackage{xcolor}
\usepackage{hyperref}
\usepackage{comment}
\newcommand\notsotiny{\@setfontsize\notsotiny{7.2}{8.2}}
\newcounter{takeaway}
\newcommand{\takeawaylabel}[1]{\def\thetakeawaylabel{#1}\label{takeaway:#1}}
\newenvironment{takeaway}[1][]{\refstepcounter{takeaway}\par\medskip
   \noindent \textbf{Takeaway~\thetakeaway. #1} \rmfamily}{\medskip\takeawaylabel{\thetakeaway}}
\begin{document}
\title{Mining for sustainability in cloud architecture among the discussions of software practitioners: building a dataset\thanks{\scriptsize This work is partly funded by the project SustainableCloud (project number OCENW.M20.243) of the research program Open Competition Domain Science by the Dutch Research Council (NWO).}}
\author{Sahar Ahmadisakha\orcidID{0000-0002-1485-4512} \and
Vasilios Andrikopoulos\orcidID{0000-0001-7937-0247}}
\authorrunning{S. Ahmadisakha}
\institute{University of Groningen, Netherlands\\
\email{\{s.ahmadisakha, v.andrikopoulos\}@rug.nl}
}
\maketitle              
\begin{abstract}
The adoption of cloud computing is steadily increasing in designing and implementing software systems, thus it becomes imperative to consider the sustainability implications of these processes. While there has already been some academic research on this topic, there is a lack of perspective from practitioners. To bridge this gap, we utilize software repository mining techniques to examine 192 discussions among practitioners on the Software Engineering forum of the StackExchange platform, aiming to build an annotated dataset containing cloud architectural discussions and to understand the current discussion on sustainability in cloud architecture. To identify these discussions, we first put together a list of terms indicating sustainability as the topic. Our initial findings indicate practitioners mainly focus on design aspects—analysis, synthesis, and implementation—while avoiding complex activities like evaluation and maintenance. Technical sustainability is emphasized, while the economic dimension has the most discussions exclusively focused on it. This contrasts with previous academic literature, which highlighted environmental sustainability.

\keywords{cloud computing \and sustainability dimensions \and mining software repository \and software architecture.}
\end{abstract}
\section{Introduction}
Although sustainability in software engineering has a longstanding history, its integration into software architecture is a relatively recent development~\cite{lago-2015}. This shift has been profoundly influenced by the rise of cloud computing, which refers to the provision of internet-accessible computing resources~\cite{mell2011nist}. Notably, this emergence has foregrounded the environmental sustainability aspect, primarily due to the substantial energy consumption of data centers, which serve as the primary hosts for cloud infrastructures~\cite{andrikopoulos2021}.

Extensive research endeavors have already been undertaken at the intersection of software architecture and sustainability, aiming to address sustainability~\cite{lago-2019, venters-2018, venters2023sustainable} or evaluate it~\cite{koziolek2011sustainability} and understand it from the practitioners' perspective~\cite{condori2020action, condori2018characterizing}. However, there have been limited efforts to explore the implications of cloud computing on this intersection like~\cite{AHMADISAKHA2024107459} investigating all dimensions from an academic view and~\cite{vos2022architectural} focusing on environmental dimension from the practitioners' perspective.

The \textbf{objective} of this study is therefore to ascertain the perspectives of software practitioners regarding sustainability dimensions within cloud architectural discussions. The dimensions include \emph{technical}, \emph{economic}, \emph{environmental}, and \emph{social}, concerning software longevity, capital preservation, natural resources conservation, and community continuity, respectively~\cite{lago-2015}. We aim to identify how sustainability is recognized within these discussions and to what extent its various dimensions are addressed. 
To achieve our objective, we utilize software repository mining techniques to analyze discussions among practitioners on the \textit{Software Engineering (SE)} forum of StackExchange, and build a dataset that contains cloud architectural discussions along with their associated sustainability dimensions. Despite the presence of previous studies~\cite{tian2019developers, bi2021mining, de2023characterizing} that explore architectural knowledge in \textit{Stack Overflow}, we specifically focus on SE and do not include other forums such as \textit{Stack Overflow} or \textit{Code Review}, as they yielded a small useful sample in our pilot query. We believe that the selected forum is appropriate enough since SE is dedicated to software engineering and it is more likely to contain discussions on architectural topics.

The primary \textbf{contributions} of this study are therefore twofold: \textit{Firstly}, the establishment of an \textit{initial set of terms} for conducting searches related to sustainability in corpora. This set could be used for both systematic surveys of the literature, and for mining studies in the future. \textit{Secondly}, the creation of a validated and annotated \textit{dataset of practitioners' discussions} related to the various sustainability dimensions in conjunction with cloud architecting, identifying specific quality requirements and architecting phases in them. This annotated dataset provides a valuable resource for further academic research on the topic. It also offers insights to researchers and practitioners, highlighting the current state and elicitation methods of sustainability dimensions in architectural discussions.

The paper continues with an expanded study design in Section~\ref{sec:design}, followed by results in Section~\ref{sec:results}, and discussion in Section~\ref{sec:discussion}. Threats to validity are addressed in Section~\ref{sec:ttv}, with related works in Section~\ref{sec:related}. Conclusions and future work are presented in Section~\ref{sec:conclusion}, with a statement on data availability in Section~\ref{sec:data}.

\section{Study Design}
\label{sec:design}
As the goal of this work is to establish the practitioners' perspective on the topic of sustainability in architecting cloud-based software systems, we opt to utilize the established checklist provided by the ACM SIGSOFT Empirical Standards for Software Engineering~\cite{ralph2020empirical} as the methodological backbone to conduct a mining software repositories (MSR) study, as outlined in the following.

\subsection{Research Aims and Questions}
\noindent The objective of this study using the Goal-Question-Metric~\cite{van2002goal} formulation is:

\begin{hangparas}{2em}{0}
\rightskip=2em
\textit{\textbf{Analyze} the experiences and opinions of software practitioners in discussion forums \textbf{for the purpose of} understanding and building a dataset from the practitioners' perspective about sustainability, \textbf{with respect to} the associated dimensions \textbf{from the point of view} of the software practitioners \textbf{in the context of} architecting cloud-based software systems.}
\end{hangparas}

\noindent 
This investigation is conducted through the lens of \textbf{quality requirements} by examining discussions in software repositories that we call ``data points''. The stated goal leads to two research questions:\\

\begin{hangparas}{3.1em}{1}
\textbf{RQ1: }\textit{Which architectural phases are considered in the discussions?}
\end{hangparas}
\vspace{0.5\baselineskip}
\noindent Since SE is not a forum specific to software architecture and is generally about software engineering, and in order to ensure a comprehensive analysis based on architectural discussions, it is crucial to identify only the relevant discussions. For this purpose, we use data points that encompass \emph{at least one phase of the architecting life cycle} so-called \emph{analysis, synthesis, evaluation, implementation, and maintenance}~\cite{tang2010comparative}, to be able to elicit the necessary architectural knowledge such as quality requirements. By means of identifying these phases, we also aim to understand the evolution of cloud architecture discussions over time.
\vspace{0.5\baselineskip}

\begin{hangparas}{3.1em}{1}
\textbf{RQ2: }\textit{What is the state of the discussions with regard to the sustainability dimensions?}
\end{hangparas}
\vspace{0.5\baselineskip}
\noindent At the same time, we also aim to comprehend the status of sustainability considerations within software repositories through discussions among software practitioners. This endeavor is essential for understanding the communication of sustainability aspects within these repositories and enables us to compare those findings against those previously extracted from the literature in existing work~\cite{AHMADISAKHA2024107459}.

\subsection{Dataset building}
This section of the study design encompasses three steps that we further elaborate on, namely: \textit{Repository selection}, \textit{Query building}, and \textit{Data point selection}. The overall process is summarized later in the paper in Fig.~\ref{fig:queryfiltering}.
\subsubsection{Step 1: Repository selection}
For our study's objectives, we already selected Q\&A platforms, notably Stack Exchange (SEx), as it offer discussions on software engineering topics. Among the SEx forums, we focused on the SE forum. We chose this because SEx lacks a specific platform dedicated to discussing issues or ideas surrounding software systems architecture and because it is a forum for software engineering discussions. Moreover, recent studies~\cite{de2022developers} have indicated that Q\&A websites, such as Stack Exchange (and its forums), are admitted by practitioners as the most valuable source of architectural information.
\subsubsection{Query terms identification}
To collect data from the above-mentioned repository, we utilized a query through the SEx data explorer (SEDE)\footnote{\url{https://data.stackexchange.com/} [Accessed: 18 March 2024]}, same as~\cite{tahir2020large}. Formulating this query involves selecting appropriate terms tailored to our study. Due to the goal of this work, our query revolves around three key concepts: \emph{software architecture}, \emph{cloud computing}, and \emph{sustainability}. We incorporated multiple terms for each concept to maximize the inclusion of relevant data points. In the following, we discuss the rationale behind the selection of the query terms as they appear summarized in Table~\ref{tab:terms}. The full list of terms and the query itself are available in the replication package of our study (see Section~\ref{sec:data}).\\ 

\begin{table*}[t!]
\caption{Below are the terms included in our mining query. We removed the ``*'' sign from certain terms like \texttt{architect*} or \texttt{efficien*}, as it is handled by ``\%'' in the actual query. Please note that only a portion of the sustainability terms are included.}
\label{tab:terms}
\centering
\renewcommand{\arraystretch}{1.2}
\scriptsize
\begin{tabularx}{\linewidth}{>{\raggedright\arraybackslash}p{0.17\linewidth} *{5}{>{\raggedright\arraybackslash}X}}
\toprule
\textbf{Category} & \multicolumn{5}{l}{\textbf{Query Terms}} \\
\midrule
\textbf{Architecture} & \texttt{architect} & \texttt{design} & \texttt{tactic} & \texttt{pattern} & \texttt{best practice} \\
                     & \texttt{model} & \texttt{principle} & \texttt{analysis} & \texttt{synthesis} & \texttt{maintenance} \\
                     & \texttt{evaluation} & \texttt{implementation} & & & \\
\midrule
\textbf{Cloud}        & \texttt{AWS} & \texttt{EC2} & \texttt{S3} & \texttt{Aurora} & \texttt{Azure} \\
                      & \texttt{Azure DevOps} & \texttt{Azure Virtual} & \texttt{Azure Blob} & \texttt{Google Cloud} & \texttt{GCP}  \\
                      & \texttt{Google Compute} & \texttt{GCE} & \texttt{Google App} & \texttt{GAE} & \texttt{GCS}  \\
\midrule
\textbf{Sustainability} & \texttt{sustain} & \texttt{technic} & \texttt{econom} & \texttt{soci} & \texttt{environment} \\
                        & \texttt{cost} & \texttt{effort} & \texttt{efficien} & \texttt{evol} & \texttt{resource} \\
                        & \texttt{performance} & \texttt{continu} & \texttt{complexity} & \texttt{consum} & \texttt{...} \\
\bottomrule
\end{tabularx}
\end{table*}
\noindent\textit{\textbf{Query terms related to software architecture:}} For this we adopt the \texttt{architect*} term and choose the main software architecture's terms from~\cite{bass-2021}. We also add the architecting phases defined in~\cite{tang2010comparative} as other terms to this part of the query. For more information about the query terms see Table~\ref{tab:terms}.\\

\noindent\textit{\textbf{Query terms related to cloud computing:}} We include the term \texttt{cloud} and the names of the top three cloud providers based on~\href{https://info.flexera.com/CM-REPORT-State-of-the-Cloud?lead_source=Organic%20Search}{Flexera's State of the Cloud Report, 2024}: Amazon Web Services, Microsoft Azure, Google Cloud Platform, along with their abbreviations (AWS, Azure, GCP), and their three most commonly used cloud services in the query. Sources for top services for~\href{https://mindmajix.com/top-aws-services}{AWS},~\href{https://blog.clairvoyantsoft.com/microsoft-azure-and-its-most-used-services-15580f2105d5}{Azure}, and~\href{https://medium.com/@shiva.yarlagadda89/top-10-gcp-services-3b33f77dcb4a}{GCP} are in the hidden links. 

\noindent\textit{\textbf{Query terms related to sustainability: }} The most challenging aspect of constructing the search query lies in selecting the appropriate set of query terms for sustainability. To our knowledge, there is no predefined list of keywords specifically tailored for searching sustainability topics that cover all its dimensions. Moreover, our pilot query with the term \texttt{sustain*}, similar to most secondary studies, yielded zero results. This suggests the need for additional search terms for sustainability. 
Existing studies~\cite{koziolek2011sustainability, garcia2018interactions} often rely on terms like ``green'', ``environment*'', or ``sustainab*'', but previous research highlights limitations in capturing sustainability content solely through these terms~\cite {AHMADISAKHA2024107459}. This challenge stands as a significant burden in sustainability research. This led us to explore alternative terms for investigating sustainability topics, inspired by previous work in mining software repositories from an energy efficiency aspect~\cite{albonico2021mining}. To this effect, we initiated a three-step process (T1 to T3) to identify additional terms beyond basic sustainability terms, as outlined in Fig.~\ref{fig:term-extraction}; more specifically: 

\begin{figure*}[t]
\centering
\includegraphics[width=.75\textwidth]{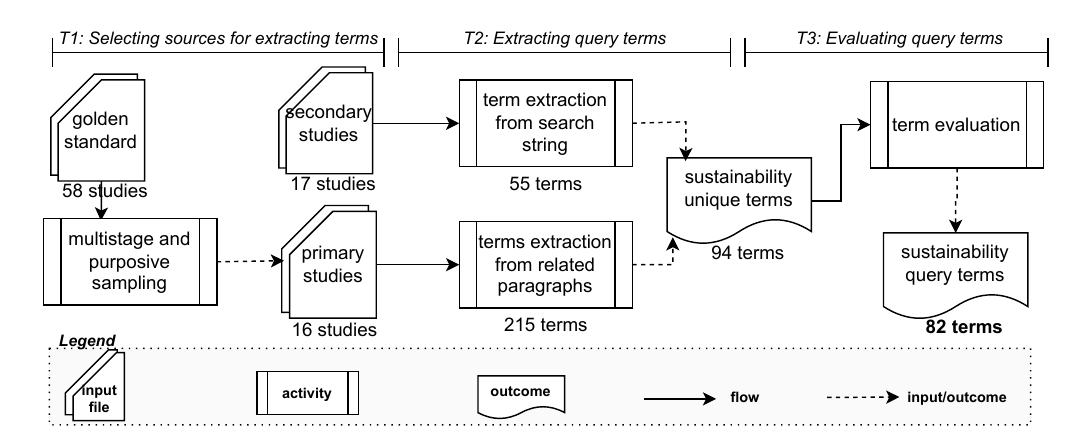}
\caption{The process of extracting sustainability-related terms for further query building step. \textbf{82} such terms are extracted.}
\label{fig:term-extraction}
\end{figure*}

\noindent\textit{\textbf{T1. Sources for extracting sustainability query terms}}

\noindent We initiate our term pool by identifying related terms from two sources: primary, and secondary literature studies. A list of 17 \textit{secondary studies} (see replication package in Section~\ref{sec:data} for the full list) was chosen based on their coverage of sustainability in software engineering and architecture research, and on whether they were conducted in a systematic manner. For \textit{primary studies}, we use the dataset of systematic mapping study of Andrikopoulos et al.~\cite{andrikopoulos-2022} in two ways: as an additional source of terms come from primary studies, and as a \emph{gold standard} for assessing whether the terms we selected are suitable for retrieving sustainability-related works (T3, below).

We start by sampling the set of primary studies in~\cite{andrikopoulos-2022}. To sample the dataset, we cluster studies based on their addressed sustainability dimensions. We employ various techniques, including \textit{multistage} sampling guided by methodologies outlined in~\cite{baltes2022sampling}, and \textit{purposive} sampling, integrating both probability and non-probability sampling methods. Given the variability in the number of sustainability dimensions addressed by primary studies (ranging from 1 to 4), we prioritize studies covering a single dimension. This allows us to select 3 primary studies per dimension, as we believe they offer more tailored and specific terms. Down this road, we faced challenges with the economic and social dimensions. No primary studies were exclusively dedicated to the economic dimension in~\cite{andrikopoulos-2022}. To address this, multistage sampling was used. We selected a \textit{cluster} including studies addressing 2 dimensions, one of which relates to the economic dimension, and chose 3 primary studies \textit{randomly} from them in the cluster. Regarding the social dimension, only 1 study focused solely on it, which we included in our selection. We also identified 3 studies addressing the social dimension along with another dimension and \textit{randomly} selected 2 of them. For the technical and environmental dimensions, we \textit{randomly} chose 3 primary studies for each among the studies that address solely the technical or environmental dimension.

Finally, we bolstered our initial selection of 12 primary studies by adding 4 more studies through \textit{purposive sampling}. These additional studies covered all four sustainability dimensions. We included them because comprehensive coverage of all dimensions enhances the focus on sustainability, increasing the likelihood of identifying relevant terms.\\
\begin{figure*}[t]
\centering
\includegraphics[width=.85\textwidth]{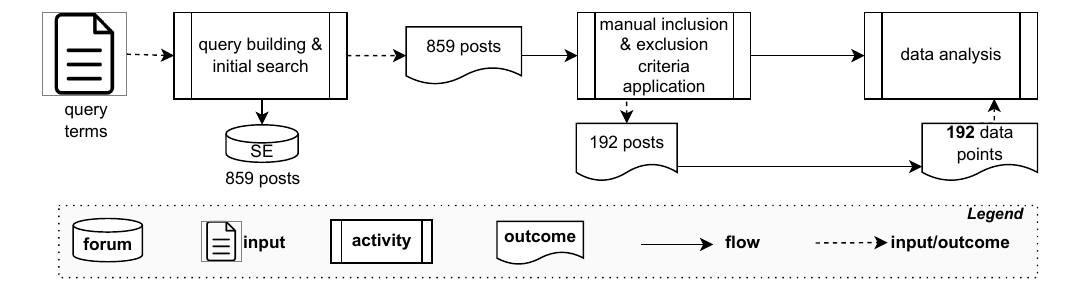}
\caption{The process of extracting and filtering data points from SEDE to analysis. \textbf{859} data points are extracted among which \textbf{192} data points are accepted.}
\label{fig:queryfiltering}
\end{figure*}

\noindent\textit{\textbf{T2. Extracting sustainability query terms}}

\noindent Upon completing the process of identifying relevant studies for extracting sustain-ability-related terms, we proceed to extract query terms from these sources. In this process, before doing any extraction, we initially select the terms \texttt{sustain*, technic*, econom*, environment*}, and \texttt{soci*} as \textbf{basic terms} of sustainability and its dimensions. From the 17 secondary studies, we extract sustainability-related terms from their search strings. For the 16 selected primary studies, we thoroughly surveyed the text to identify paragraphs discussing \textbf{basic terms} and gathered any technical, economic, social, or environmental terms mentioned. We also examine the text for sustainability-related concepts, such as ``responsible resource use for green computing,'' aligning with the environmental dimension.
During term extraction, we highlighted relevant lines from papers, ensuring each term was only extracted once per document. We then normalized the terms, reducing 270 initial terms to \textbf{94 unique terms}, and mapped both initial and normalized terms to sustainability dimensions, as detailed in the replication package.\\

\noindent\textit{\textbf{T3. Evaluating sustainability query terms}}

\noindent As a way of evaluating the resulting set of terms, we check whether we can retrieve with them all 58 primary studies from our gold standard dataset. We search all 94 terms across these studies (excluding their references sections). By using all 94 terms, we indeed can retrieve all 58 primary studies. However, we could see that while some terms like \texttt{time, sustain*, change, user, quality, maint*}, and \texttt{value} seem promising in retrieving 100-88 percent of the studies, except for \texttt{sustain*}, they lack specificity and may vary in meaning depending on context. Therefore, we opt to eliminate them and search for a \textit{minimal set} within the remaining terms.
A minimal set comprises the fewest terms needed to retrieve all primary studies. To find one, we sort terms by their coverage of primary studies in descending order of appearance frequency and sequentially add these terms to the minimal set until all studies from~\cite{andrikopoulos-2022} can be retrieved using the set. A minimal set of \textbf{82 terms}, available in the replication package, is constructed through this process.

\subsubsection{Step 2: Query building}
After identifying the appropriate terms for the query, we build the query as follows:
\begin{quote}
    \texttt{([software architecture-related terms in OR]) AND ([cloud computing-
    related terms in OR]) AND ([sustainability-related terms in OR])}
\end{quote}
\noindent using the terms in Table~\ref{tab:terms} for software architecture and cloud computing, and the 82 terms from T3 in the previous step for sustainability, as shown in Fig.~\ref{fig:queryfiltering}.

\subsubsection{Step 3: Data point selection}
After executing the query we proceed by selecting the necessary columns from the \texttt{Posts} table in the SEDE results. We retain 5 out of the 23 available columns, namely: \texttt{Id} which is a unique identifier for each post; \texttt{PostTypeId} which is the type of post (e.g., question, answer); \texttt{CreationDate} which is date and time when the post was created; \texttt{Score} which is the total score of the post, contains the net sum of upvotes and downvotes; and \texttt{Body} which is the main content of the post, including text, code snippets, and other relevant information. 
Thereafter, we apply our inclusion and exclusion criteria, outlined in Table~\ref{tab:IncExc}, to filter relevant data points. Any post meeting all inclusion criteria and none of the exclusion criteria progresses to the next stage which is analysis. After applying our criteria to the \textbf{initial 859 data points}, we focus on the remaining \textbf{192 accepted data points} to address our research questions. To note: we accept only 22\% of posts. Most rejections (64\%) are due to false positives (E1). Only 10\% are not architecturally relevant (I2), and the remaining 4\% involve posts with negative scores or inappropriate IDs (I1, E2). With only 10\% of discussions being architecturally irrelevant, the Software Engineering forum on SEx \textit{can be} ideal for architectural discussions.
\begin{table*}[t!]
\caption{Inclusion and exclusion criteria. I$N$ stands for inclusion criterion number and E$N$ stands for exclusion criterion number.}
\label{tab:IncExc}
\centering
\renewcommand{\arraystretch}{1.2}
\scriptsize
 \begin{tabularx}{\linewidth}{>{\centering\arraybackslash}p{0.05\linewidth} >{\raggedright\arraybackslash}p{0.35\linewidth} >{\raggedright\arraybackslash}p{0.6\linewidth}}
\toprule
 \textbf{ID} & \textbf{Criteria} & \textbf{Explanation}\\
 \toprule
 I1 & Zero or positive \texttt{Score} data point & We want to include data point that has not been given negative scores from other practitioners. \\
 I2 & Data point is about an architectural discussion & We want to keep data points that necessarily have an architectural discussion.\\
 E1 & False positive data point & We have to reject the data point if it includes no data or perception on cloud computing or is out of the topic.\\
 E2 & \texttt{PostTypeId} > 2 & In Posts tables, we want to keep data points that are either question or answer. (\texttt{PostTypeId} = 1 or 2 respectively)\\
  \bottomrule
 \end{tabularx}
\end{table*}

\subsection{Data extraction and initial analysis}
We \textit{manually} extract and organize the data in a spreadsheet, mainly focusing on the main discussions in the posts' \texttt{Body} column.

\subsubsection{RQ1} To address this research question, we read each data point to determine which architecting phase it pertains to based on the activities reported in~\cite{tang2010comparative}. This evaluation (and the next one for RQ2) is primarily conducted manually by one researcher, with random checks by a second researcher to ensure consistency.
\subsubsection{RQ2} To address our research question, we identify the sustainability dimensions discussed within each data point. Since data points do not directly address sustainability dimensions, we first extract the quality requirements (QR) mentioned in the entire post. We then map these QRs to sustainability dimensions using results in~\cite{AHMADISAKHA2024107459} and the sustainability quality model outlined in~\cite{condori2018characterizing} as a guide (the mapping table is available in the replication package). However, 72 data points do not explicitly mention any QRs. For these, we assign appropriate QRs by exercising our judgment (indicated by \texttt{coded-qr} in the dataset, together with the post's relevant part). As an illustration, consider a part of data point \texttt{ID~122098} which implicitly talks about \texttt{fault tolerance} and \texttt{scalability}: 
\begin{quote}
\footnotesize
\textit{...To me the key point is the cloud is a set of cheap, unreliable resources. \textbf{Your solution has to be built to keep running when the resources fail, and have a scalable architecture that can utilize additional resources as they are added}.}
\end{quote}
As an illustration of how QR to sustainability dimension mapping happens, consider data point \texttt{ID~432796}:
\begin{quote}
\footnotesize
\textit{..This doesn't feel quite right, am I missing a \textbf{security}/privacy measure here?
I'm thinking about pushing this app soon, potentially marketing it but, at first, simply handing the APK to family and friends for wider use and testing of the beta - I'll have access to their most personal memories. Is it a matter of \textbf{trust} or is there some way I can implement a certificate or something somehow for?..}
\end{quote}
This example demonstrates the presence of two quality requirements: \texttt{security} and \texttt{trust}. As per the mappings in~\cite{condori2018characterizing, AHMADISAKHA2024107459}, security contributes to the technical and social dimensions, while trust contributes to the technical, economic, and social dimensions. Consequently, this data point reflects aspects of all dimensions except for the environmental one. Moreover, it indicates that the discussion primarily focuses on technical and social aspects, as evident from the quotation.
\\
\\
A part of data point \texttt{ID~360022} as another example of this mapping is presented:
\begin{quote}
\footnotesize
\textit{...First Decision: Self Hosted vs the Cloud:
The cloud allows you to use AWS S3 or whatever equivalent blob storage for your cloud provider. \textbf{This solution only charges you for the storage you use, and cloud blob storage provides both the scale and performance needed to scale as your application grows}...}
\end{quote}
This quotation discusses a design decision for selecting the hosting infrastructure for an application, highlighting the cloud's ability to scale with growth and charge only for used resources. This underscores the economic importance of \texttt{scalability} in resource utilization. The economic contribution of scalability is also noted by~\cite{condori2018characterizing}, and we incorporated this in the mapping.
\section{Preliminary Results and Findings}
\label{sec:results}
\begin{figure*}[t]
\centering
\includegraphics[width=\textwidth, height=0.85\textheight, keepaspectratio]{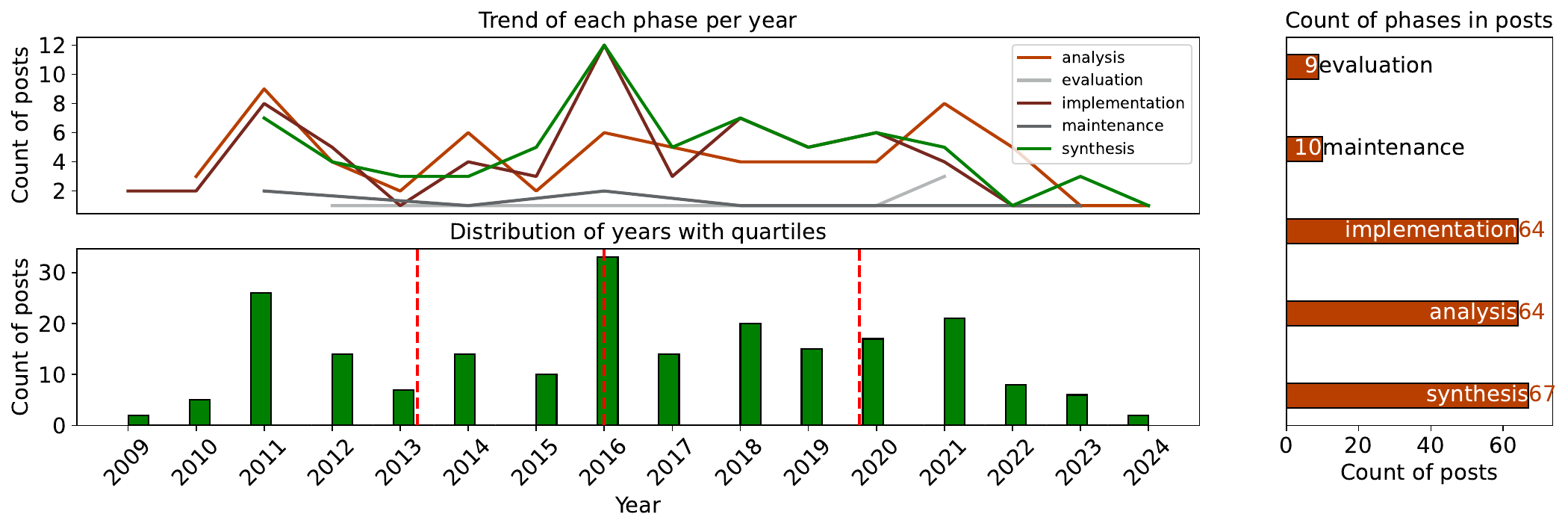}
\caption{From top-left--\textit{Upper plot}: architecting phases addressed along the years. \textit{Lower plot}: Cloud architecture discussions across years. \textit{Vertical plot}: Architecting phases counts that are addressed in the SE posts discussions}
\label{fig:year-phase}
\end{figure*}
In Fig.~\ref{fig:year-phase}, we illustrate the distribution of discussions over 16 years. While the lower plot suggests a binomial distribution, the data from the first half occurred within nearly eight years, centered around 2016, which marks the second quartile (Q2) of the data. This implies a slightly higher frequency of architectural discussions during the early years of cloud computing emergence.
\subsection{RQ1: architectural discussion coverage}
\label{subsec:res-phases}
%
Based on the architectural phases identified during the data extraction process, one observation pertains to the distribution of discussions across architectural phases emerges, as illustrated in Fig.~\ref{fig:year-phase}. Notably, there appears to be a balanced representation of the analysis, synthesis, and implementation phases. However, the relatively limited presence of maintenance and evaluation discussions may arise from various factors. Regarding maintenance, it might be due to the comprehensive support provided by cloud service providers. In the context of cloud computing, tasks such as data center management, operating system updates, security patching, load balancing, resource monitoring, data backup, and disaster recovery are typically handled by the cloud infrastructure, potentially reducing the need for extensive discussion on maintenance activities.

In terms of the evaluation phase, it appears that assessing cloud architectures may not be the primary focus of interest. Rather, there is a greater emphasis on discussing how to design and deploy architectures for cloud environments. This suggests that practitioners might be primarily engaged in discussing the appropriate way for building and migrating applications to operate effectively within cloud environments. Consequently, having frequent discussions about analysis, synthesis, and implementation phases \textit{may} be attributed to this ongoing exploration and advancement of cloud-based architecture practices.

Another point can be made by examining the architectural phases over the years from Fig~\ref{fig:year-phase}. The upper plot in the figure indicates similar behaviors between the analysis and implementation phases before Q2 (which is around 2016), with a subsequent divergence after Q2 where the implementation phase continues with synthesis. We speculate that the fact that implementation phases have evolved with analysis somehow near 2016 and then continued with synthesis might be related to the first official demonstration of \href{https://aws.amazon.com/blogs/architecture/announcing-updates-to-the-aws-well-architected-framework/}{AWS's well-architected framework} that emerges at 2016, as supported by the biggest cloud provider. This framework may have given architects more means to do implementation and build on top of the cloud providers.
\subsection{RQ2: State of the sustainability discussions}
\label{subsec:res-dims}
%
%
%
By extracting quality requirements from the discussions and mapping them to sustainability dimensions, it becomes apparent that a discussion may encompass one or multiple sustainability dimensions.
\begin{figure*}[t]
\centering
\includegraphics[width=0.8\textwidth, height=0.65\textheight, keepaspectratio]{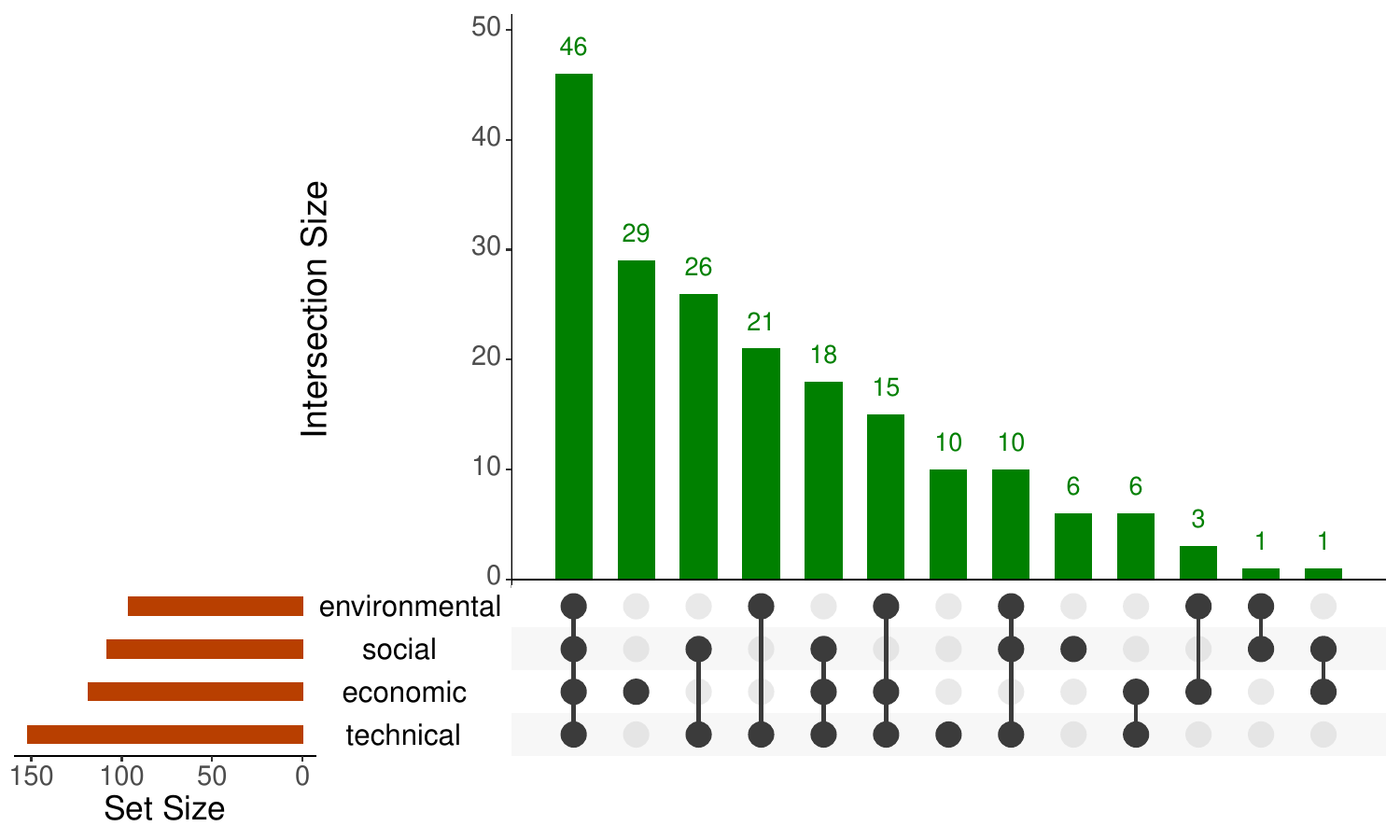}
\caption{Discussed sustainability dimensions along with their intersections.}
\label{fig:upset-dims}
\end{figure*}
%
After doing the mapping for all data points, the resulting analysis reveals the distribution of data points containing any of the sustainability dimensions, as depicted in Fig.~\ref{fig:upset-dims}. Notably, the technical dimension emerges as the most commonly mapped, which aligns with the tendency of practitioners to mainly address and tackle technical aspects.

Figure~\ref{fig:upset-dims} also illustrates the intersections among various dimensions that we call combinations. Notably, these intersections closely correspond to those revealed in academic literature in~\cite{AHMADISAKHA2024107459} except for the combination of social, economic, and environmental dimensions (\texttt{soc-eco-env}), and solo environmental (\texttt{env}) which are not detected in the practitioners' discussions. In the current study, we identify four additional combinations (\texttt{env-soc}, \texttt{env-soc-tec}, \texttt{tec-soc-eco-env}), and (\texttt{soc}) which are not detected in~\cite{AHMADISAKHA2024107459}. We share 11 combinations with that reported in~\cite{AHMADISAKHA2024107459}.  
Among the newly found intersections, one notable example is the combination of technical, social, economic, and environmental dimensions (\texttt{tec-soc-eco-env}), which surprisingly emerges as the most frequent intersection. This is one of the combinations that is not found at all in~\cite{AHMADISAKHA2024107459}. Upon closer examination of the figure, we observe that after the \texttt{tec-soc-eco-env} combination, the economic dimension emerges as the second most popular. Next frequent combinations predominantly involve the technical-social (\texttt{tec-soc}) and technical-environmental (\texttt{tec-env}) dimensions. This observation suggests a potential correlation between the technical dimension and both social and environmental dimensions.
{\begin{takeaway}
    \label{takeaway:counts}
        \noindent\textcolor[HTML]{3F00FF}{\textit{The economic dimension is primarily addressed individually, indicating that it not only accounts for the highest number of data points solely focused on this dimension, but also makes a substantial contribution to the majority of the discussions. This is evident from its position as the second dimension in terms of set size in Figure~\ref{fig:upset-dims}. This underscores the significance of economic considerations in the discussions among practitioners.}}
    \end{takeaway}    
    }
%
%
%

\begin{figure*}[t]
\centering
\includegraphics[width=0.7\textwidth, height=0.6\textheight, keepaspectratio]{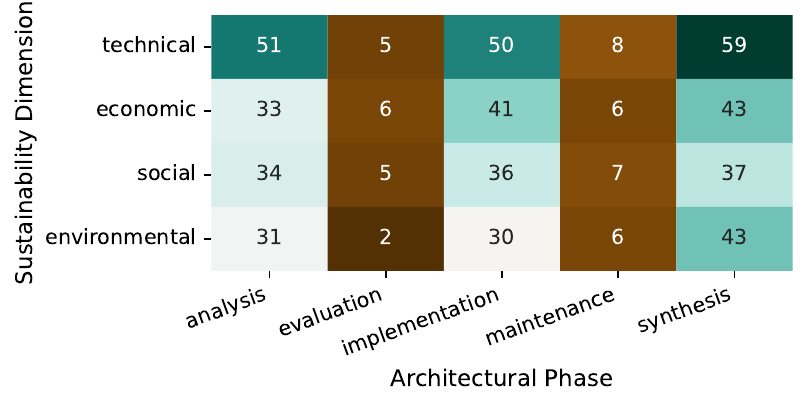}
\caption{The co-occurrence of sustainability dimensions with architectural phases.}
\label{fig:heatmap-phase-dim}
\end{figure*}

Based on the data we have collected thus far from the dataset and initial analysis, the final aspect we wish to present concerns the intersection of architectural phases and sustainability dimensions, as illustrated in Fig.~\ref{fig:heatmap-phase-dim}. The figure indicates that all sustainability dimensions are addressed across all architectural phases; however, the frequency of their co-occurrence in the evaluation and maintenance phases is notably low, for all dimensions. As is high for analysis, synthesis, and implementation across all dimensions. In summary, sustainability dimensions do not appear to significantly impact the extent to which each architectural phase is communicated. This observation may be attributed to the specific forum selected or the context of cloud computing.
\section{Discussion}
\label{sec:discussion}
Based on our findings, we would like to identify some points of discussion that require further consideration.

To start off, a list of terms for searching about sustainability is presented in this study. This list corresponds to the shortage of suitable terms for researching sustainability, as highlighted by~\cite{AHMADISAKHA2024107459}. \emph{Our approach, which entails identifying sustainability-related query terms, does not claim to cover every potential study or capture all sustainability-related terms. Instead, it represents an initial effort to identify terms for further sustainability research which is sufficient to fulfill our needs in this study.} Further efforts to both augment the list and further examine it are part of our future work. The question, however, arises: if the list is adequate for sustainability research, why are most data points rejected? The answer lies in the inclusion and exclusion criteria we have set for this study. We do not evaluate data points based on their direct connection to sustainability, as this is achieved through QR mapping. The posts, as discussed and presented in Section~\ref{sec:design}, are mostly rejected because they do not provide a cloud architectural discussion. These rejections stem from various issues, such as inappropriate usage of cloud terms, the inclusion of cloud terms just within links, mentioning only a cloud service without substantive discussion, and providing definitions of cloud computing, etc. See this sample that \texttt{cloud} term is used irrelevantly:
\begin{quote} \footnotesize \textit{ Time to step down from the \textbf{clouds}. Composites actually describe the meat of the code. I'm talking about classes, methods, objects, functions, prototypes...} 
\end{quote}

\emph{The findings pertaining to the most addressed dimension in Section~\ref{subsec:res-dims} diverge from those reported in the academic literature~\cite{AHMADISAKHA2024107459} in the same context of cloud architecting. Previous research highlights the notable significance of the environmental dimension in academia, with the technical dimension ranking third.} This divergence could be attributed to the specific repository under analysis or may underscore differences in how sustainability is perceived and prioritized by practitioners compared to academics. It is important to note that the results presented here are comparable only to those of~\cite{AHMADISAKHA2024107459}, as it is the sole study sharing a similar context with the current investigation, specifically in terms of analyzing \textit{cloud architectural} discussions. We would like to mention one of the takeaways:
{\begin{takeaway}
    \label{takeaway:shares-2}
        \noindent\textcolor[HTML]{3F00FF}{\textit{In the context of cloud computing, academia tends to prioritize environmental sustainability, while practitioners exhibit a stronger focus on technical sustainability. This indicates potential variations in the prioritization and communication of sustainability dimensions between academic discourse and practical implementation within cloud architecting.}}
    \end{takeaway}    
    }

Another interesting point to add to the discussion is that in 2021, the AWS well-architected framework (that we first mentioned in Section~\ref{subsec:res-phases}) integrated a sustainability pillar, with a strong emphasis on the environmental dimension of cloud-based software. Almost all of the major cloud providers, including \href{https://azure.microsoft.com/en-us/solutions/cloud-enablement/well-architected#reliability}{Microsoft Azure} and \href{https://cloud.google.com/architecture/framework/}{Google Cloud Platform}, have developed similar frameworks. However, we do not observe any increase in the number of discussions related to sustainability in the years following 2021, and definitely no increase in the environmental sustainability dimensions. \emph{This may indicate a potential disinterest by the practitioners in the topic of environmental sustainability.} Future work to clarify this disparity is required.

Finally, we believe that our approach of extracting architectural discussions in general --- that is, without considering sustainability or cloud computing as part of the picture --- from the SEx platform has been effective, with a minimal rejection rate of 10\%. \emph{However, considering the comprehensive architecture centers and sustainability best practices offered by major cloud providers like AWS and Azure, leveraging their community forums (e.g.\ \href{https://aws.amazon.com/forums/}{AWS}, \href{https://azure.microsoft.com/en-us/support/community}{Azure}) instead of the general-purpose SE forum at SEx may offer greater advantages.} The challenge lies in the lack of straightforward methods to obtain forum data through downloads or queries from these forums, prompting us to continue mining data from the SEx for the time being.

\section{Threats to Validity}
\label{sec:ttv}
\noindent\textbf{External validity:} Utilizing only one forum of SEx, meaning SE, creates a potential external validity threat. However, the most relevant forum in this Q\&A platform to our study are those, and there are recent studies in our field like~\cite{tahir2020large} that similarly SE as one of three relevant forums of SEx. We excluded Stack Overflow (SO) and Code Reviews (CodeR) due to their emphasis on programming issues, which are not in our interest. Despite expecting higher results from SO and similar results from CodeR in comparison with SE (considering their whole number of questions and answers to be higher and equal with SE), our tailored query yielded fewer data entries from them (SO and CodeR), indicating their potential irrelevancy to our topic.
Access to the Stack Exchange Data Explorer (SEDE) was rechecked on March 18th, 2024, to update the study's results. For reproducibility, the complete~\href{https://figshare.com/s/64519d3bb4edaf047174}{replication package} of this research is made available online, as discussed in Section~\ref{sec:data}. However, a potential concern arises regarding the data acquisition method, since similar to prior recent studies like~\cite{tahir2020large} we query the SEDE which may not retrieve all available data due to caching and request capping. To address this concern, our query was executed multiple times using different browsers, users, and timestamps to minimize potential data loss.

\noindent\textbf{Internal validity:} We employ a predefined set of terms for querying data points, a method proven effective in energy-related software mining (e.g.,~\cite{cruz2019energy}, \cite{pinto2014mining}). To mitigate potential drawbacks, such as false positives, \textit{we manually analyzed all data points to prevent any false positive inclusions}. To address potential false negatives from the initial query, we utilized over 90 different queries, systematically adding each sustainability-related term. The number of query results becomes constant around the term \texttt{flexibility}, indicating saturation. However, to ensure comprehensive coverage, we included the remaining terms in the final query. Details regarding the saturation point and results per query are provided in the replication package. While we have just created the initial query terms for sustainability, we cannot guarantee that we have eliminated all possible false negatives. However, by using various queries and employing terms from relevant literature, we aimed to minimize this risk.

\section{Related Works}
\label{sec:related}
In the realm of mining software repositories, extensive research has explored various aspects of software development, yet a notable gap persists: the investigation of sustainability dimensions. While numerous studies have delved into energy-related topics in robotics and Android applications, as well as broader aspects of the software development life cycle, sustainability dimensions remain relatively unexplored. For example, Cruz et al.~\cite{cruz2019energy} examined the impact of energy-oriented code changes on the maintainability of Android applications, providing insights into the intersection of energy efficiency and application development. Similarly, Albonico et al.~\cite{albonico2021mining} and Malavolta et al.~\cite{malavolta2021mining} have recently investigated energy-related practices within robotics software in related repositories including StackOverflow, addressing gaps in knowledge and offering comprehensive analyses of energy efficiency in robot operating systems.

In a different track, efforts by Pinto et al~\cite{pinto2014mining} on StackOverflow and Moura et al.~\cite{moura2015mining} explored software energy consumption concerns among application programmers, providing valuable insights into energy-aware strategies for improving energy efficiency in real-world applications. A bit on social aspects and architectural concepts, in parallel, the study by Tizard et al.~\cite{tizard2019requirement} discusses software project failures attributed to a lack of user feedback and missed requirements, proposing mining product forums as a solution to address this challenge. Koetter et al.~\cite{koetter2019assessing} delve into the intersection of product quality requirements and social aspects within software projects. They analyze commit history to extract success and failure factors, providing recommendations to enhance project quality, including maintainability. While these studies touch upon architectural, social, or energy-related aspects that may intersect with sustainability, none explicitly address sustainability dimensions in the context of software repositories, particularly within cloud architectural discussions. Recognizing this gap, we embarked on the journey \textit{to construct a dataset} focused explicitly on sustainability dimensions, laying the groundwork for future works in this crucial area of software architecting.

\section{Conclusion and Future Work}
\label{sec:conclusion}
We analyzed data points from the SEx repository (specifically from the SE forum) \textit{to understand and build a dataset from practitioners' perspectives on sustainability in cloud architectural discussions}, considering the growing importance of sustainability in such systems. Initially, we extracted architectural phases and quality requirements from each data point (post) and then annotated them with the respective sustainability dimension(s) by means of quality requirements mapping to sustainability dimensions. We build a dataset consisting of posts containing cloud architectural discussions, together with the extracted QRs and associated sustainability dimensions. Our initial findings suggest that the SE forum holds promise for architectural discussions and that there is a notable emphasis on technical and economic sustainability. Furthermore, practitioners engaged in cloud architecting are still in the process of discussing how to design, implement, and deploy applications to the cloud. Although we found a suitable forum for architectural discussions (SE), we recognize the need for additional sites and repositories such as those discussed in Section~\ref{sec:discussion}. Additionally, we propose revisiting the sustainability annotation process based on quality requirement mappings through consensus building to enhance the robustness of our findings.
\section{Data Availability}
\label{sec:data}
Data associated with this research is publicly available online in the replication package provided at the~\href{https://figshare.com/s/64519d3bb4edaf047174}{link}.
\bibliographystyle{splncs04}
\bibliography{REF}

\clearpage

\end{document}